Research article

**Open Access**

# Zone center phonons of the orthorhombic RMnO$_3$ (R = Pr, Eu, Tb, Dy, Ho) perovskites

HC Gupta*[1] and Upendra Tripathi[2]

Address: [1]Physics Department, Indian Institute of Technology, Hauz Khas, New Delhi, India and [2]Physics Department, Amity University, Noida, UttarPradesh, India

Email: HC Gupta* - hcgupta@physics.iitd.ernet.in; Upendra Tripathi - t_upendra1974@yahoo.co.in

* Corresponding author

## Abstract

A short range force constant model (SRFCM) has been applied for the first time to investigate the phonons in RMnO$_3$ (R = Pr, Eu, Tb, Dy, Ho) perovskites in their orthorhombic phase. The calculations with 17 stretching and bending force constants provide good agreement for the observed Raman frequencies. The infrared frequencies have been assigned for the first time.

**PACS Codes:** 36.20.Ng, 33.20.Fb, 34.20.Cf

## Introduction

Until recently the RMnO$_3$ perovskites (R = rare earth elements) have been the object of research mainly as parent materials of mixed valence manganites exhibiting colossal magnetoresistivity (CMR) [1-4]. In the past few years, however, there is an increased interest in the complex relationships among the lattice distortions, magnetism, dielectric, and transport properties of undoped RMnO$_3$ [5-10]. All RMnO$_3$ perovskites show a distortion of MnO$_6$ octahedra due to orbital ordering characteristic of the John-Teller effect of Mn$^{3+}$ cations [11-15]. An investigation of infrared and Raman frequencies will be quite useful in describing the details of such properties. Practically, very limited information is available on the infrared and Raman scattering of orthorhombic RMnO$_3$. Martin Carron *et al.* [11] studied the behavior of Raman phonons through the transition from static to dynamic Jahn-Teller order in stoichiometric RMnO$_3$ samples (R = La, Pr, Y). Also Martin Carron *et al.* [12] studied orthorhombic RMnO$_3$ (R = Pr, Nd, Eu, Tb, Dy, Ho) manganites for their Raman phonons as a function of the rare earth ions and temperature. They had assigned only some of the Raman modes. They correlated the frequencies of three most intense modes of orthorhombic samples, with some structural parameters such as Mn-O bond distances, octahedral tilt angle and Jahn-Teller distortion. Further rationalization of the Raman spectra of orthorhombic $RMnO_3$ (R = Pr, Nd, Tb, Ho, Er) and different phases of Ca- or Sr- doped $RMnO_3$ compounds as well as cation deficient $RMnO_3$ were made by Martin Carron *et al.* [13]. Their assignment of the peaks related to octahedral tilt were in good agreement with the other authors but the assignment of peak to an antisymmetric stretching associated with the Jahn-Teller distortion was doubtful. Wang Wei-Ran *et al.* [14] measured Raman active phonons in orthorhombic $RMnO_3$ (R = La, Pr, Nd, Sm) compounds and they also assigned three main Raman peaks. Recently, the polarized Raman spectra of orthorhombic $RMnO_3$ (R = Pr, Nd, Eu, Gd, Tb, Dy, Ho) series at room temperature were studied by Iliev *et al.* [15] where they had assigned the observed frequencies to nine Raman modes. Their study shows that the variations of lattice distortions with radius of rare earth atoms affect significantly both the phonon frequencies and the shape of some of Raman modes. To our knowledge, the theoretical investigations of phonons, using the normal coordinate analysis in the orthorhombic $NdMnO_3$ has first been made by Gupta *et al.* [16].

In the present study, the theoretical investigations of phonons in the orthorhombic $RMnO_3$ have been made using the normal coordinate analysis. It has been observed that a total of 17 inter-atomic force constants, which include 8 bending force constants, are enough to obtain a good agreement between theory and experiment for the Raman frequencies. The assignments of infrared frequencies along with their corresponding eigen vectors observing the atomic displacements in the respective vectors have been made for the first time. There is always some scope of more precise infrared experiments to verify these theoretical values.

**Theory**

The structure of stoichiometric $RMnO_3$ shown in Fig. 1, described at room temperature by the Pbnm space group (Z = 4), can be considered as orthorhombically distorted superstructure of ideal perovskites. In the Pbnm structure the atoms occupy four non equivalent atomic sites of them only the Mn site is a center of symmetry [17]. The distortion of the orthorhombic perovskites characterized by the tilting angle of the $MnO_6$ octahedra progressively increases from Pr to Er due to simple steric factors. Additionally, all of the perovskites show a distortion of the $MnO_6$ octahedra due to orbital ordering characteristic of the Jahn-Teller of the $Mn^{3+}$ cations. Structural data of $EuMnO_3$ is very recent because of its high neutron absorption and they are perfectly correlated with the other members of $RMnO_3$ series [18].

The total number of irreducible representations for $RMnO_3$ are

$$= 7A_g + 7B_{1g} + 5B_{2g} + 5B_{3g} + 8A_u + 8B_{1u} + 10B_{2u} + 10B_{3u}$$

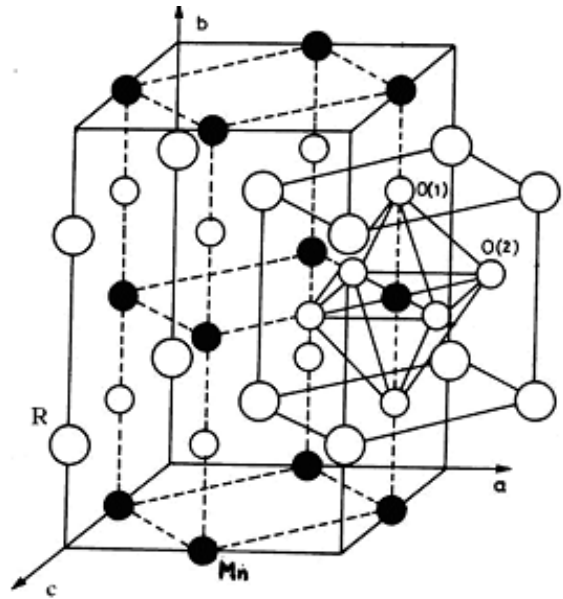

**Figure 1**
The structure of Orthorhombic RMnO$_3$ (R = Pr, Nd, Eu, Gd, Tb, Dy, Ho) compounds at room temperature, belonging to Pbnm space group. The structure has four formulae unit with R atoms, Mn atoms and O atoms (O1 and O2).

There are four Raman active species, $A_g$, $B_{1g}$, $B_{2g}$ and $B_{2g}$, three infrared active species $B_{1u}$, $B_{2u}$ and $B_{3u}$ and inactive specie $A_u$.

In the present paper, an attempt has been made to study the zone center phonons in RMnO$_3$ (R = Pr, Eu, Tb, Dy, Ho) for the first time using SRFCM. We have used nine valence force constants $K_1$(Mn-O2), $K_2$(Mn-O1), $K_3$(Mn-O2), $K_4$(R-O1), $K_5$(R-O2), $K_6$(R-O1), $K_7$(R-O2), $K_8$(R-O1), $K_9$(R-O2); and eight bending force constants $H_1$(O1-R-O1), $H_2$(O1-R-O1), $H_3$(O1-R-O1), $H_4$(O1-R-O2), $H_5$(O1-R-O2), $H_6$(O1-R-O2), $H_7$(O2-R-O2) and $H_8$(O2-R-O2) at various interatomic distances and angles as shown in Table 1(only for PrMnO$_3$).

**Table 1: Force constant, Coordination number, Inter-atomic Distances (Å) and Angles (deg) and Force constant values (N/cm) for Orthorhombic PrMnO$_3$**

| Force constant | $K_1$ | $K_2$ | $K_3$ | $K_4$ | $K_5$ | $K_6$ | $K_7$ | $K_8$ | $K_9$ | $H_1$ | $H_2$ | $H_3$ | $H_4$ | $H_5$ | $H_6$ | $H_7$ | $H_8$ |
|---|---|---|---|---|---|---|---|---|---|---|---|---|---|---|---|---|---|
| Coord. Number. | 8 | 8 | 8 | 4 | 8 | 4 | 8 | 4 | 8 | 8 | 8 | 4 | 4 | 8 | 8 | 7 | 8 |
| Distance/Angle | 1.91 | 1.95 | 2.21 | 2.36 | 2.40 | 2.48 | 2.62 | 3.17 | 3.52 | 89 | 67 | 110 | 90 | 56 | 66 | 160 | 120 |
| Force constant values | 0.597 | 0.535 | 0.950 | 0.456 | 0.019 | 0.311 | 0.382 | 0.335 | 0.598 | 0.432 | 0.413 | 0.404 | 0.373 | 0.338 | 0.329 | 0.136 | 0.022 |

## Results and Discussions

A systematic variation in the most of the force constants is seen throughout the series. It was interesting to observe that although, the interatomic distances for $K_1$ and $K_3$ between Mn and O2 atoms remain nearly unchanged from Pr to Ho but the force constant exhibited a uniform increase. This behaviour can be related to the increase in distortion of $MnO_6$ octahedra. Further, as shown in Table 1 the force constant $K_3$ (0.950 N/cm) is quite large when compared with the similar force constant obtained in studies of $NdNiO_3$ [19] and $NdGaO_3$ [20] (0.620 N/cm). A similar kind of behaviour of large force constant between Mn and O2 atoms was observed in pyrochlore manganates [21]. This may be one of the possible reasons of associated CMR properties of manganese compounds. To account for a drastic change in resistivity and a low critical temperature in such materials, it should be noted that the double exchange model must be combined with the effect of the Jahn-Teller distortion of $MnO_6$ octahedra [22]. This effect promotes carrier localization and dresses charge carriers *via* cloud of phonons. It is in this respect where the large interatomic force between Mn and O2 atoms plays an important role, being a part of the distortion of the $MnO_6$ octahedra. The force constants between R and O1 atoms, $K_4$ and $K_6$ increase with decrease of R-O1 distance almost uniformly throughout the series. The force constant $K_8$ (R-O1) changes by a small amount as the R-O1 distance also shows the similar behavior. The force constants $K_5$, $K_7$ and $K_9$ also show a uniform increase. Although force constant $K_5$ is very small but K$_9$ shows comparatively a large value. The bending force constants H$_1$-H$_4$ show a very small change in force constant values while H$_7$ and H$_8$ exhibit uniformly increasing values.

Table 2: *Observed [15] and Calculated Raman Wave Numbers ($cm^{-1}$) for Orthorhombic $RMnO_3$ (R = Pr, Eu, Tb, Dy, Ho)

| Modes | *Pr | Pr | *Eu | Eu | *Tb | Tb | *Dy | Dy | *Ho | Ho |
|---|---|---|---|---|---|---|---|---|---|---|
| $A_g$ | 491 | 491 | 501 | 501 | 509 | 509 | 513 | 513 | 520 | 520 |
|  | 462 | 462 | 479 | 479 | 489 | 489 | 492 | 492 | 499 | 499 |
|  |  | 386 |  | 392 |  | 402 |  | 412 |  | 408 |
|  | 324 | 324 | 361 | 361 | 378 | 378 | 386 | 386 | 395 | 395 |
|  | 232 | 232 |  | 270 |  | 269 |  | 272 | 288 | 288 |
|  |  | 206 |  | 205 |  | 211 |  | 213 |  | 210 |
|  |  | 64 |  | 67 |  | 79 |  | 79 |  | 77 |
| $B_{1g}$ | 607 | 607 | 610 | 610 | 612 | 612 | 614 | 614 | 615 | 615 |
|  | 496 | 496 | 518 | 518 | 528 | 528 | 534 | 534 | 537 | 537 |
|  |  | 486 |  | 499 |  | 501 |  | 501 |  | 503 |
|  | 445 | 445 | 465 | 465 | 474 | 474 | 478 | 478 | 481 | 481 |
|  | 312 | 312 | 324 | 324 | 331 | 331 | 336 | 336 |  | 340 |
|  |  | 114 |  | 122 |  | 127 |  | 129 |  | 129 |
|  |  | 84 |  | 91 |  | 96 |  | 97 |  | 97 |
| $B_{2g}$ |  | 627 |  | 611 |  | 621 |  | 624 |  | 617 |
|  |  | 492 |  | 511 |  | 519 |  | 521 |  | 529 |
|  |  | 432 |  | 463 |  | 469 |  | 476 |  | 482 |
|  |  | 283 |  | 295 |  | 302 |  | 306 |  | 309 |
|  |  | 125 |  | 131 |  | 134 |  | 134 |  | 135 |
| $B_{3g}$ |  | 537 |  | 521 |  | 545 |  | 553 |  | 546 |
|  |  | 400 |  | 429 |  | 432 |  | 432 |  | 454 |
|  |  | 305 |  | 367 |  | 381 |  | 390 |  | 402 |
|  |  | 239 |  | 266 |  | 270 |  | 274 |  | 286 |
|  |  | 123 |  | 124 |  | 127 |  | 127 |  | 125 |

**Table 3: Calculated Infrared Wave Numbers (cm$^{-1}$) for Orthorhombic RMnO$_3$ (R = Pr, Eu, Tb, Dy, Ho)**

| Modes | Pr | Eu | Tb | Dy | Ho |
|---|---|---|---|---|---|
| B$_{1u}$ | 608 | 611 | 612 | 614 | 617 |
|  | 569 | 581 | 581 | 580 | 582 |
|  | 485 | 492 | 509 | 514 | 516 |
|  | 303 | 323 | 328 | 332 | 338 |
|  | 205 | 213 | 214 | 214 | 223 |
|  | 141 | 152 | 158 | 159 | 161 |
|  | 133 | 135 | 142 | 144 | 143 |
|  | 0 | 0 | 0 | 0 | 0 |
| B$_{2u}$ | 614 | 612 | 617 | 620 | 620 |
|  | 571 | 582 | 582 | 580 | 580 |
|  | 467 | 494 | 498 | 500 | 511 |
|  | 389 | 395 | 406 | 417 | 410 |
|  | 290 | 304 | 309 | 312 | 318 |
|  | 223 | 229 | 232 | 234 | 235 |
|  | 201 | 206 | 208 | 208 | 213 |
|  | 177 | 176 | 180 | 179 | 178 |
|  | 132 | 142 | 148 | 148 | 149 |
|  | 0 | 0 | 0 | 0 | 0 |
| B$_{3u}$ | 535 | 538 | 551 | 558 | 562 |
|  | 484 | 505 | 515 | 519 | 522 |
|  | 431 | 458 | 463 | 465 | 474 |
|  | 343 | 384 | 398 | 406 | 419 |
|  | 315 | 320 | 318 | 316 | 315 |
|  | 244 | 268 | 272 | 277 | 289 |
|  | 181 | 181 | 185 | 184 | 184 |
|  | 131 | 137 | 143 | 144 | 143 |
|  | 106 | 115 | 118 | 120 | 122 |
|  | 0 | 0 | 0 | 0 | 0 |

The calculated Raman frequencies in Table 2 agreed satisfactorily with the observed values [15]. The assignment of infrared frequencies as shown in Table 3 has been done for the first time. Still a precise experimental analysis of infrared frequencies is needed to verify the results of present calculations. The potential energy distribution (PED) for most of the force constant is found to be almost similar throughout the series. The PED showed that high wave numbers are dominated by stretching force constants involving Mn and O atoms and bending force constants having R and O atoms. Therefore, the symmetric stretching of the basal oxygens of the octahedra, around 610 cm$^{-1}$ (B$_{1g}$ symmetry); the asymmetric stretching at about 490 cm$^{-1}$ (A$_g$ symmetry) associated with the Jahn-Teller distortion is expected. The A$_g$ mode (324 cm$^{-1}$ - 395 cm$^{-1}$) showing a drastic increase in frequency is purely a stretching mode dominated by K$_9$ (R-O2). Most of the lower wave number modes have a convincing influence by R-O bending and stretching force constants. For all the compounds of the orthorhombic RMnO$_3$ series, we calculated the eigen vectors representing the displacements of various atoms. It was observed that for larger wave numbers, the displacement of O atoms is important whereas for smaller wave numbers, the displacement of R atoms dominates as given in Table 4 and Table 5 only for PrMnO$_3$. Vibrations of several atoms are involved in some middle order modes.

**Table 4: Calculated Raman Wave Numbers (cm$^{-1}$) of PrMnO$_3$ along with their Eigen-vector Lengths representing Atomic Displacements for various Atoms**

| Modes | Wave-numbers | Pr | Pr | O1 | O1 | O2 | O2 | O2 |
|---|---|---|---|---|---|---|---|---|
| A$_g$ | 491 | 0.04 | 0.26 | -0.08 | -0.46 | 0.69 | -0.43 | 0.21 |
| | 462 | 0.05 | 0.16 | -0.24 | 0.53 | 0.60 | 0.52 | -0.02 |
| | 386 | 0.05 | 0.05 | 0.96 | 0.05 | 0.20 | 0.15 | 0.01 |
| | 324 | 0.12 | -0.15 | -0.06 | -0.66 | 0.09 | 0.54 | -0.47 |
| | 232 | -0.30 | 0.21 | -0.03 | -0.27 | -0.18 | 0.47 | 0.74 |
| | 206 | 0.91 | -0.16 | -0.04 | -0.01 | -0.07 | 0.06 | 0.36 |
| | 64 | 0.24 | 0.90 | 0.00 | -0.01 | -0.27 | 0.01 | -0.25 |
| B$_{1g}$ | 607 | -0.03 | 0.10 | -0.06 | 0.96 | -0.05 | 0.09 | 0.24 |
| | 496 | 0.33 | 0.05 | 0.78 | 0.06 | -0.49 | -0.13 | -0.10 |
| | 486 | 0.05 | 0.00 | 0.10 | -0.08 | -0.07 | 0.99 | -0.03 |
| | 445 | 0.07 | -0.17 | 0.51 | 0.06 | 0.82 | 0.01 | 0.14 |
| | 312 | -0.27 | 0.34 | 0.15 | -0.25 | -0.12 | 0.00 | 0.84 |
| | 114 | 0.28 | 0.90 | -0.05 | -0.01 | 0.24 | 0.00 | -0.23 |
| | 84 | 0.85 | -0.18 | -0.30 | -0.07 | 0.01 | -0.01 | 0.38 |
| B$_{2g}$ | 627 | 0.01 | | 0.90 | | 0.13 | 0.37 | 0.20 |
| | 493 | 0.08 | | -0.20 | | 0.95 | 0.20 | -0.06 |
| | 432 | -0.17 | | -0.27 | | -0.25 | 0.88 | -0.25 |
| | 283 | 0.09 | | -0.28 | | -0.05 | 0.19 | 0.94 |
| | 125 | 0.98 | | -0.01 | | -0.12 | 0.11 | -0.13 |
| B$_{3g}$ | 537 | 0.06 | | 0.59 | | 0.41 | 0.68 | -0.08 |
| | 400 | 0.04 | | -0.23 | | -0.69 | 0.64 | 0.25 |
| | 305 | 0.21 | | -0.73 | | 0.41 | 0.32 | -0.40 |
| | 239 | 0.28 | | -0.18 | | 0.39 | 0.00 | 0.86 |
| | 123 | 0.93 | | 0.19 | | -0.20 | -0.14 | -0.18 |

**Table 5: Calculated Infrared Wave Numbers (cm⁻¹) of PrMnO$_3$ along with their Eigen-vector Lengths representing Atomic Displacements for various Atoms**

| Modes | Wave-numbers | Mn | Mn | Mn | Pr | Pr | O1 | O1 | O2 | O2 | O2 |
|---|---|---|---|---|---|---|---|---|---|---|---|
| B$_{1u}$ | 606 | -0.02 | -0.02 | 0.01 | -0.10 | | 0.95 | | 0.11 | -0.06 | 0.25 |
| | 569 | 0.02 | -0.53 | 0.03 | 0.01 | | -0.10 | | 0.84 | -0.04 | -0.02 |
| | 485 | 0.08 | -0.05 | -0.27 | -0.07 | | 0.09 | | 0.03 | 0.94 | -0.16 |
| | 303 | -0.19 | 0.01 | -0.17 | -0.35 | | -0.26 | | 0.02 | 0.12 | 0.86 |
| | 205 | 0.95 | 0.01 | 0.20 | -0.20 | | -0.05 | | -0.01 | 0.00 | 0.15 |
| | 141 | -0.24 | -0.25 | 0.76 | -0.48 | | -0.01 | | -0.17 | 0.17 | -0.12 |
| | 133 | -0.07 | 0.81 | 0.23 | -0.14 | | -0.04 | | 0.50 | 0.08 | -0.07 |
| | 0 | 0.00 | 0.00 | 0.47 | 0.76 | | 0.00 | | 0.00 | 0.26 | 0.36 |
| B$_{2u}$ | 614 | -0.01 | -0.03 | 0.00 | 0.07 | 0.04 | 0.02 | 0.93 | -0.25 | 0.14 | 0.00 |
| | 571 | 0.02 | -0.53 | 0.02 | 0.06 | -0.06 | 0.00 | -0.14 | -0.01 | 0.83 | -0.04 |
| | 467 | -0.01 | -0.01 | -0.03 | -0.03 | -0.30 | -0.22 | 0.21 | 0.88 | 0.03 | 0.18 |
| | 389 | -0.04 | 0.00 | 0.10 | -0.04 | -0.07 | 0.96 | 0.01 | 0.19 | 0.01 | 0.12 |
| | 290 | -0.27 | 0.01 | -0.08 | -0.10 | 0.03 | -0.08 | -0.24 | -0.15 | 0.02 | 0.91 |
| | 223 | -0.14 | 0.42 | 0.00 | 0.81 | -0.32 | 0.01 | -0.08 | -0.06 | 0.17 | 0.02 |
| | 201 | 0.93 | 0.10 | 0.19 | 0.06 | -0.06 | -0.01 | -0.07 | -0.05 | 0.03 | 0.28 |
| | 177 | -0.20 | 0.04 | 0.97 | -0.07 | -0.03 | -0.12 | 0.00 | -0.01 | 0.01 | 0.00 |
| | 132 | -0.02 | 0.56 | -0.08 | -0.56 | -0.46 | -0.01 | 0.01 | -0.17 | 0.35 | -0.10 |
| | 0 | 0.00 | 0.47 | 0.00 | 0.00 | 0.76 | 0.00 | 0.00 | 0.26 | 0.36 | 0.00 |
| B$_{3u}$ | 535 | -0.25 | 0.07 | -0.04 | -0.21 | -0.07 | 0.27 | 0.59 | -0.49 | 0.46 | -0.09 |
| | 484 | 0.15 | -0.05 | 0.10 | -0.28 | 0.03 | 0.85 | -0.21 | -0.10 | -0.33 | 0.09 |
| | 431 | -0.21 | 0.10 | -0.04 | -0.04 | 0.19 | 0.30 | 0.16 | 0.81 | 0.35 | -0.04 |
| | 343 | -0.04 | -0.53 | -0.03 | -0.20 | 0.22 | -0.14 | 0.57 | 0.15 | -0.45 | 0.25 |
| | 315 | 0.05 | 0.82 | -0.02 | -0.12 | 0.13 | -0.09 | 0.25 | 0.01 | -0.31 | 0.35 |
| | 244 | -0.25 | -0.14 | -0.01 | 0.21 | -0.19 | 0.07 | -0.16 | -0.05 | 0.24 | 0.86 |
| | 181 | -0.26 | 0.03 | 0.95 | 0.14 | 0.05 | -0.03 | 0.06 | 0.00 | -0.07 | -0.05 |
| | 131 | -0.06 | -0.03 | -0.06 | 0.12 | 0.93 | 0.00 | -0.19 | -0.24 | 0.15 | 0.06 |
| | 106 | 0.71 | -0.06 | 0.29 | -0.40 | 0.05 | -0.15 | 0.02 | 0.06 | 0.41 | 0.21 |
| | 0 | 0.47 | 0.00 | 0.00 | 0.76 | 0.00 | 0.26 | 0.36 | 0.00 | 0.00 | 0.00 |